\documentclass[12pt]{article}

\usepackage[english]{babel}
\usepackage[cp1251]{inputenc}
\usepackage[T1]{fontenc}

\pdfoutput=1
\usepackage{makeidx}
\usepackage{amssymb}
\usepackage{amsfonts}
\usepackage{amsmath}
\usepackage{graphicx}
\usepackage{setspace}
\usepackage{authblk}
\usepackage{units}
\usepackage{appendix}
\usepackage{xparse}
\usepackage{cite}
\usepackage{hyperref}
\usepackage[left=2.1cm,top=2.5cm,right=2.1cm]{geometry}

\begin{document}

\title{Coherent states of an accelerated particle}
\author{A. I. Breev$^{1}$\thanks{
		breev@mail.tsu.ru}, D. M. Gitman$^{2,3}$\thanks{
		dmitrygitman@hotmail.com}, Paulo A. Derolle$^{3}$\thanks{paderolle@usp.br}
	, \\
	$^{1}$\small{Department of Physics, Tomsk State University, \\
		Lenin ave. 36, 634050 Tomsk, Russia;}\\
	$^{2}$ P.N. Lebedev Physical Institute, \\
	53 Leninskiy ave., 119991 Moscow, Russia.\\
	$^{3}$ Institute of Physics, University of S\~{a}o Paulo, \\
	Rua do Mat\~{a}o, 1371, CEP 05508-090, S\~{a}o Paulo, SP, Brazil.}

\maketitle

\abstract{
	We construct generalized coherent states (GCS) of a massive accelerated particle. This example is an important step in studying coherent states (CS) for systems with an unbounded motion and a continuous spectrum. First, we represent quantum states of the accelerated particle both known and new ones obtained by us using the method of non-commutative integration of linear differential equations. A complete set of non-stationary states for the accelerated particle is obtained. This set is expressed via elementary functions and is characterized by a continuous real parameter $\eta$, which corresponds to the initial momentum of the particle. A connection is obtained between these solutions and stationary states, which are determined by the Airy function.
	We solved the problem of constructing GCS, in particular, semiclassical states describing the accelerated particle, within the framework of the consistent method of integrals of motion. We have found different representations, coordinate one and in a Fock space, analyzing in detail all the parameters entering in these representations. 
	
}

\section{Introduction\label{S1}}

Coherent states (CS) play an important role in modern quantum theory as
states that provide a natural relation between quantum mechanical and
classical descriptions \cite{Schro26,Glaub63,Glaub63b,Glaub63c}. They have a number of useful
properties and as a consequence a wide range of applications, e.g. in
semiclassical description of quantum systems, in quantization theory, in
condensed matter physics, in radiation theory, in quantum computations and
so on, see, e.g. Refs.                                                             \cite{CSQT72,CSQT72b,CSQT72c,CSQT72d,CSQT72e,MalMa,DodMa87,DodMa03,105,Bergeron24}. Despite the
fact that there exist a great number of publications devoted to constructing
CS of different systems, an universal definition of CS and a constructive
scheme of their constructing for arbitrary physical system is not known. In
this relation, it should be noted that CS were first introduced and studied
in detail for systems with bounded motion and discrete spectrum like
harmonic oscillator, charged particle in a magnetic field and so on.
Formally the problem of constructing CS for systems with quadratic
Hamiltonians of the general form was solved in works by Dodonov and Man'ko,
using Malkin and Man'ko integral of motion method, see cited above Refs..
However, it should be noted that sometimes to extract appropriate sets of CS
from the general results is not a simple task. Even for the simplest and                      
physically important system as a free particle, the problem of CS
construction was, in fact, solved relatively recently, in Ref. \cite{303}
following, in fact, the integral of motion method, and its special version
proposed in Ref. \cite{311}. We believe that this situation is explained by
the fact that the free particle represents an unbounded motion with the
continuous energy spectrum and a generalization of the initial (Glauber)
scheme in constructing CS of a harmonic oscillator was not so obvious in
this case. In fact, a formal application of the integral of motion method to
systems with unbounded motion results in constructing the so-called
generalized coherent states (GCS). In Ref. \cite{303} the attention was paid
on the fact that among families of formally constructed GCS there may exist
both semiclassical states and quantum states which do not describe any
semiclassical motion at all. Fixing special parameters which arise in the
construction of GCS can be distinguished from physical consideration she
families of semiclassical states, in particular CS as well as squeezed
states.

In this article, we, using the integral of motion method, construct GCS of a
massive accelerated particle. This example is a next important step in
studying CS for systems with an unbounded motion and a continuous spectrum.
Besides of its physical importance there is a didactic advantage of using
accelerated particle CS in teaching of quantum mechanics. On this example,
we once again demonstrate the existence of GCS that describe both
semiclassical and purely quantum motions. In this regard, it should be said
that the problem of constructing semiclassical states describing an
unbounded motion with some time-dependent Hamiltonians, was considered in
relatively recent works based on various approaches; see e.g. Refs. \cite%
{Nagiev2018,Sazon23,Pereira,Sazon24}. This interest stresses the importance
of the problem under consideration. We consider constructing GCS, in
particular, semiclassical states describing an accelerated particle, within
the framework of the above-mentioned consistent method of integrals of
motion mentioned above. In Sec. (\ref{S2}), we study quantum states of the
accelerated particle both known and new ones obtained by us using the method
of non-commutative integration of linear differential equations.\ In Sec. (%
\ref{S3}), we construct GCS of the accelerated particle, in different
representations analyzing in detail all the parameters entering in the
constructions. We prove the corresponding completeness and orthogonality
relations. Standard deviations and conditions of the semi-classicality are
discussed in Sec. (\ref{S4}). Here we define the so-called CS and a class of
CS that can be identified with semiclassical states. In the Conclusion (\ref%
{S5}), we tried to list technical and physical results obtained in this
article that are important in our opinion.

\section{Some exact solutions of the Schr\"{o}dinger equation\label{S2}}

One of the adequate approaches to the quantum description of the rectilinear
accelerated motion of a nonrelativistic particle seems to be the
consideration of the motion of the particle in an uniform external field.
Let we have one-dimensional motion along the $x$-axis, and let $F$ be the
constant force acting on the particle. The potential energy $U$ can be taken
as $U=-Fx$, such that the corresponding Hamiltonian reads ($\hat{p}%
_{x}=-i\hbar d/dx$):%
\begin{equation}
	\hat{H}_{x}=\frac{\hat{p}_{x}^{2}}{2m}-Fx=-\frac{\hbar ^{2}}{2m}\frac{d^{2}}{%
		dx^{2}}-Fx\ .  \label{1.1}
\end{equation}%
In particular, if the particle has an electric charge $e$ such a force can
be caused by an electric field of the intensity $\mathcal{E}$, i.e., 
$F=e\mathcal{E}$. For example, for the particle of mass $m$ near the Earth's
surface, where the gravitational field is almost constant, it is acted upon
by a constant force $F=-mg$, where $g$ is the gravity of Earth.

Below, we recall known stationary solutions of the Schr\"{o}dinger equation
with the Hamiltonian (\ref{1.1}) and construct new non-stationary solutions
of the corresponding time-dependent Schr\"{o}dinger equation using the
method of non-commutative integration of linear differential equations \cite%
{nc1,nc2,nc3}.

\subsection{Stationary states\label{S2.1}}

In the coordinate representation, stationary states $\psi _{E}\left(
x\right) $ satisfy the Schr\"{o}dinger equation $\hat{H}_{x}\psi _{E}\left(
x\right) =E\psi _{E}\left( x\right) $,%
\begin{equation}
	\frac{d^{2}\psi _{E}\left( x\right) }{dx^{2}}+\left( \frac{2m}{\hbar ^{2}}%
	\right)\left( E+Fx\right) \psi _{E}\left( x\right) =0\ .  \label{1.2}
\end{equation}%
In the potential field under consideration the energy levels form a
continuous nondegenerate spectrum, $+\infty >E>-\infty $. The corresponding
motion is finite towards $x=-\infty $ and infinite towards $x=+\infty $.
Introducing a dimensionless variable 
\begin{equation}
	\xi =\left( x+\frac{E}{F}\right) \left( \frac{2mF}{\hbar ^{2}}\right) ^{1/3},
	\label{1.3}
\end{equation}%
one can reduce Eq. (\ref{1.2}) to the form $\psi ^{\prime \prime }\left( \xi
\right) +\xi \psi \left( \xi \right) =0$. A solution of the latter equation,
which is finite for all $x$, reads (see Ref. \cite{LanLi65}):%
\begin{equation}
	\psi _{E}\left( x\right) =\psi \left( \xi \right) =A\rm{Ai} \left( -\xi \right)
	,\ \mathrm{Ai} \left( \xi \right) =\frac{1}{\pi}\int_{0}^{\infty }\cos
	\left( \frac{1}{3}u^{3}+u\xi \right) du\ .  \label{1.5}
\end{equation}%
The function $\rm{Ai} \left( \xi \right) $ is the so-called Airy function, see
Ref. \cite{88}, and $A=[2m/(\hbar ^{2}\sqrt{F})]^{1/3}$ is a
normalization factor which provides the normalization of the functions $\psi
_{E}\left( x\right) $ to the delta function of the energy,%
\begin{equation}
	\int_{-\infty }^{\infty }\psi _{E}\left( x\right) \psi _{E^{\prime }}\left(
	x\right) dx=\delta \left( E^{\prime }-E\right) \ .  \label{1.6}
\end{equation}

\subsection{A complete set of nonstationary solutions\label{S2.2}}

It is convenient to introduce the dimensionless operators $q$ and $\hat{p}%
_{q}$ and the time $\tau $ as: 
\begin{equation}
	q=xl^{-1},\text{ }\hat{p}_{q}=-i\partial _{q}=\frac{l}{\hbar }\hat{p}_{x},\
	\tau =\frac{\hbar }{ml^{2}}t\ ,  \label{1.12c}
\end{equation}%
such that%
\begin{equation}
	\hat{H}_{x}=\frac{\hbar ^{2}}{ml^{2}}H_{q},\ H_{q}=\frac{\hat{p}_{q}^{2}}{2}%
	-F_{q}q,\ F_{q}=\frac{ml^{3}}{\hbar ^{2}}F_{x}\ .  \notag
\end{equation}%
In new variables (\ref{1.12c}) the evolution is described by the Schr\"{o}%
dinger equation of the form:%
\begin{align}
	& i\hbar \partial _{t}\Psi \left( x,t\right) =\hat{H}_{x}\Psi \left(
	x,t\right) \Longrightarrow \hat{S}\chi \left( q,\tau \right) =0,\ \hat{S}%
	=i\partial _{\tau }-\hat{H}_{q}\ ,  \notag \\
	& \chi \left( q,\tau \right) =\sqrt{l}\Psi \left( lq,\frac{ml^{2}}{\hbar }%
	\tau \right) ,\ \left\vert \Psi \left( x,t\right) \right\vert
	^{2}dx=\left\vert \chi \left( q,\tau \right) \right\vert ^{2}dq\ .
	\label{1.15}
\end{align}

Some solutions of the Schr\"{o}dinger equation (\ref{1.15}) can be
constructed using the method of non-commutative integration of linear
differential equations. To this end we note that symmetry operators $\hat{Y}%
_{a},\ [\hat{Y}_{a},\hat{S}]=0$, $a=1,2,3,4,$ of the Schr\"{o}dinger
equation (\ref{1.15}),%
\begin{equation*}
	\hat{Y}_{1}=-i,\ \ \hat{Y}_{2}=\partial _{q}-iF_{q}\tau ,\ \ \hat{Y}%
	_{3}=\tau \partial _{q}-\frac{i}{2}\left( F_{q}\tau ^{2}+2q\right) ,\ \ \hat{%
		Y}_{4}=\partial _{\tau }+F\hat{Y}_{3}\ ,
\end{equation*}%
form a four-dimensional solvable Lie algebra with nonzero commutation
relations%
\begin{equation}
	\lbrack \hat{Y}_{2},\hat{Y}_{3}]=\hat{Y}_{1},\ \ [\hat{Y}_{3},\hat{Y}_{4}]=-%
	\hat{Y}_{2}\ .  \label{n2}
\end{equation}%
We define an irreducible $\lambda$-representation (see Refs. \cite{nc2,nc3}%
) of the Lie algebra (\ref{n2}) by operators that act on functions of a
variable $\eta \in \left( -\infty ,\infty \right) $ and are parameterized by
two real parameters $\ j_{0}$ and $j_{1}\geq 0$,%
\begin{equation}
	\ell _{1}=ij_{0},\ \ \ell _{2}=i\left( -\eta j_{0}+j_{1}\right) ,\ \ \ell
	_{3}=\partial _{\eta },\ \ \ell _{4}=\frac{i}{2}\eta \left( \eta
	j_{0}-2j_{1}\right) \ .  \label{n2b}
\end{equation}

We will look for a complete set of solutions to the Schr\"{o}dinger
equation, which is parameterized by $\eta$, in the form:%
\begin{equation*}
	\chi (\left. q,\tau \right\vert \eta )=\int_{-\infty }^{+\infty
	}dj_{0}\int_{0}^{+\infty }dj_{1}\ \chi (\left. q,\tau \right\vert \eta
	,j_{0},j_{1})\ ,
\end{equation*}%
where functions $\chi (\left. q,\tau \right\vert \eta ,j_{0},j_{1})$ are
found as a solution to a system of first-order differential equations%
\begin{equation}
	\left( \ell _{a}+\hat{Y}_{a}\right) \chi (\left. q,\tau \right\vert \eta
	,j_{0},j_{1})=0\ .  \label{n2c}
\end{equation}%
Then the general solution of the Schr\"{o}dinger equation (\ref{1.15}) is
given by the following integral:%
\begin{equation}
	\chi \left( q,\tau \right) =\int_{-\infty }^{+\infty }C(\eta )\chi (\left.
	q,\tau \right\vert \eta )d\eta \ ,  \label{n2d}
\end{equation}
where $C(\eta)$ is an arbitrary function such that the integral in (\ref{n2d}) converges.

A solution of Eq. (\ref{n2c}), we seek in the form:%
\begin{align}
	&\chi (\left. q,\tau \right\vert \eta ,j_{0},j_{1}) =\left( 2\pi \right)
	^{-1/4}w(j_{1})\delta (j_{0}-1)\nonumber\\
	&\times \exp \left\{ -\frac{i}{2}\left[ \tau \eta
	^{2}-2\eta (\tau F_{q}-j_{1})-2\eta (q+\tau j_{1})+F_{q}\tau ^{2}(\eta
	-j_{1})+\frac{F_{q}^{2}}{3}\tau ^{3}\right] \right\} \ .  \label{n3}
\end{align}%
Substituting representation (\ref{n3}) into Eq. (\ref{1.15}), we obtain $%
j_{1}^{2}w(j_{1})=0$, which implies $w(j_{1})=\delta (j_{1})$. Taking this
into account and integrating $\chi (\left. q,\tau \right\vert \eta
,j_{0},j_{1})$ over the parameters $j_{0}\,$and $j_{1},$ we finally obtain:%
\begin{equation}
	\chi (\left. q,\tau \right\vert \eta )=\left( 2\pi \right) ^{-1/4}\exp
	\left\{ -\frac{i}{2}\left[ (\eta \tau -2q)(\eta +\tau F_{q})+\frac{F_{q}^{2}%
	}{3}\tau ^{3}\right] \right\} \ .  \label{n4}
\end{equation}%
These constructed solutions don't have a finite norm and are parameterized
by the real parameter $\eta $. However, the solutions satisfy completeness
and orthogonality relations: 
\begin{eqnarray}
	\int_{-\infty }^{+\infty }\chi ^{\ast }(\left. q,\tau \right\vert \eta )\chi
	(\left. q,\tau \right\vert \eta ^{\prime })dq &=&\delta (\eta -\eta ^{\prime
	})\ ,  \notag \\
	\int_{-\infty }^{+\infty }\chi ^{\ast }(\left. q,\tau \right\vert \eta )\chi
	(\left. q^{\prime },\tau \right\vert \eta )d\eta &=&\delta (q-q^{\prime })\ .
	\label{n5}
\end{eqnarray}

One can find a connection between solutions (\ref{n4}) and the stationary
states (\ref{1.5}) in dimensionless variables (\ref{1.12c}). Stationary
states $\chi _{\varepsilon }(q,\tau )$ in the dimensionless variables,
satisfy the equation%
\begin{equation}
	\hat{S}\chi _{\varepsilon }(q,\tau )=0,\ \ H_{q}\chi _{\varepsilon }(q,\tau
	)=\varepsilon \chi _{\varepsilon }(q,\tau )\ .  \label{s1}
\end{equation}%
We will search these solutions in the form:%
\begin{equation}
	\chi _{\varepsilon }(q,\tau )=\frac{1}{\sqrt{2\pi }}\int_{-\infty }^{+\infty
	}\ Q_{\varepsilon }^{\ast }(\eta )\chi (\left. q,\tau \right\vert \eta
	)d\eta \ .  \label{n7}
\end{equation}%
Then it follows from Eq. (\ref{s1}) that functions $Q_{\varepsilon }(\eta )$
must satisfy the equation%
\begin{equation}
	i\left( F_{q}\ell _{3}-\ell _{4}\right) Q_{\varepsilon }(\eta )=\varepsilon
	Q_{\varepsilon }(\eta )\ ,  \label{n8}
\end{equation}%
which has the following solutions:%
\begin{eqnarray}
	&&Q_{\varepsilon }(\eta )=(2\pi F_{q}^{2})^{-1/4}\exp \left[ \frac{i}{2F}%
	\left( \frac{\eta ^{3}}{3}-2\varepsilon \eta \right) \right] \ ,  \notag \\
	&&\int_{-\infty }^{+\infty }Q_{\varepsilon }^{\ast }(\eta )Q_{\varepsilon
		^{\prime }}(\eta )d\eta =\delta (\varepsilon -\varepsilon ^{\prime })\ .
	\label{n9a}
\end{eqnarray}%
Finally, with account taking of Eq. (\ref{n9a}), we obtain the explicit form
for the stationary states $\chi _{\varepsilon }(q,\tau )$:%
\begin{eqnarray}
	&&\chi _{\varepsilon }(q,\tau )=\frac{1}{\sqrt{2\pi }}\int_{-\infty
	}^{+\infty }\ Q_{\varepsilon }^{\ast }(\eta )\chi (\left. q,\tau \right\vert
	\eta )d\eta =\chi _{\varepsilon }\left( q\right) \exp \left( -i\varepsilon
	\tau \right) \ ,  \notag \\
	&&\chi _{\varepsilon }\left( q\right) =\frac{2^{1/3}}{F_{q}^{1/6}}%
	\rm{Ai} \left( -\xi \right) ,\ \ \xi =\left( q+\frac{\varepsilon }{F_{q}}%
	\right) \left( 2F_{q}\right) ^{1/3}\,,\nonumber\\
	&&\hat{H}_{q}\chi _{\varepsilon
	}\left( q\right) =\varepsilon \chi _{\varepsilon }\left( q\right) \ ,
	\label{n10}
\end{eqnarray}%
where $\rm{Ai} \left( \xi \right) $ is the Airy function; see Eq. (\ref{1.5}).
Eq. (\ref{n10}) represents the relationship between the new nonstationary
solutions (\ref{n4}) and the stationary states $\chi _{\varepsilon }(q,\tau
) $.

One can calculate the Wigner function $W(p_{q},q,\tau )$ that corresponds to
solutions (\ref{n4}),%
\begin{eqnarray}
	W(p_{q},q,\tau )&=&\frac{1}{2\pi \hbar }\int_{-\infty }^{+\infty }\chi
	^{\ast }\left( \left. q-\frac{q^{\prime }}{2},\tau \right\vert \eta \right)
	\chi \left( \left. q+\frac{q^{\prime }}{2},\tau \right\vert \eta \right)
	e^{-ip_{q}q^{\prime }}dq^{\prime }  \notag \\
	&=&\frac{1}{2\pi \hbar }\delta \left( \eta +F_{q}\tau -p_{q}\right) \ .
	\label{n11}
\end{eqnarray}%
The obtained representation reveals the physical meaning of the parameter $%
\eta$. It is the particle momentum at the initial time moment. Note that
the Wigner functions for a particle in a variable uniform field were
obtained in Ref. \cite{DodMan1984}.

Note that the constructed solutions (\ref{n4}) form a complete and
orthogonal set and are parameterized by a continuous real parameter $\eta$. Moreover, these solutions are expressed via elementary functions, which
can be useful in various applications.

\section{GCS of an accelerated particle\label{S3}}

\subsection{Integrals of motion}

First, we pass to creation and annihilation operators $\hat{a}$ and $\hat{a}%
^{\dag }$,%
\begin{equation}
	\hat{a}=\frac{q+i\hat{p}_{q}}{\sqrt{2}},\text{ }\hat{a}^{\dag }=\frac{q-i%
		\hat{p}_{q}}{\sqrt{2}},\ \left[ \hat{a},\hat{a}^{\dag }\right] =1\ .
	\label{1.13a}
\end{equation}%
In terms of these operators, the Hamiltonian $\hat{H}_{q}$ reads: 
\begin{equation}
	\hat{H}_{q}=\frac{1}{4}\left[ \hat{a}^{\dag }\hat{a}+\hat{a}\hat{a}^{\dag
	}-\left( \hat{a}^{\dag }\right) ^{2}-\hat{a}^{2}\right] -\frac{F_{q}}{\sqrt{2%
	}}\left( \hat{a}+\hat{a}^{\dag }\right) .  \label{1.14}
\end{equation}%
It can't be reduced to the first canonical form for a quadratic combination
of creation and annihilation operators, which is the oscillator-like form,
by any canonical transformation; this indicates that the spectrum of $\hat{H}$ is continuous, as shown in Ref. \cite{105}.

Let us construct an integral of motion $\hat{A}\left( \tau \right) $ linear
in the operators $\hat{q}$ and $\hat{p}_{q}$, 
\begin{equation}
	\hat{A}\left( \tau \right) =f\left( \tau \right) \hat{q}+ig\left( \tau
	\right) \hat{p}_{q}+\varphi \left( \tau \right) \ .  \label{3.1}
\end{equation}%
Here $f\left( \tau \right) $, $g\left( \tau \right) $ and $\varphi \left(
\tau \right) $ are some complex functions on the time $\tau $. For the
operator $\hat{A}\left( \tau \right) $ to be an integral of motion, it has
to commute with the equation operator $\hat{S}=i\partial _{\tau }-\hat{H}%
_{q} $, 
\begin{equation}
	\left[ \hat{S},\hat{A}\left( \tau \right) \right] =0\ .  \label{3.2}
\end{equation}%
In case if the Hamiltonian $\hat{H}_{q}\ $is self-adjoint, the adjoint
operator $\hat{A}^{\dag }\left( \tau \right) $ is an integral of motion as
well. We also demand 
\begin{equation}
	\left[ \hat{A}\left( \tau \right) ,\hat{A}^{\dag }\left( \tau \right) \right]
	=1  \label{3.7}
\end{equation}%
for $\hat{A}\left( \tau \right) $ and $\hat{A}^{\dag }\left( \tau \right) $
to be annihilation and creation operators.

To satisfy Eq. (\ref{3.2}), the functions $f\left( \tau \right) $, $g\left(
\tau \right) ,$ and $\varphi \left( \tau \right) $ have to obey the
following equations: 
\begin{equation}
	\dot{f}\left( \tau \right) =0,\ \dot{g}\left( \tau \right) -if\left( \tau
	\right) =0,\ \ \dot{\varphi}\left( \tau \right) +iF_{q}g(\tau )=0\ ,
	\label{3.3}
\end{equation}%
where derivatives in $\tau $ are denoted by dots above. The general solution
of equations (\ref{3.3}) reads: 
\begin{equation}
	f\left( \tau \right) =c_{1},\ g\left( \tau \right) =c_{2}+ic_{1}\tau ,\ \
	\varphi \left( \tau \right) =F_{q}c_{1}\frac{\tau ^{2}}{2}-iF_{q}c_{2}\tau
	+c_{3}\ ,  \label{3.4}
\end{equation}%
where $c_{1}$,$c_{2}$ and $c_{3}$ are arbitrary constants. Note that the
constant $c_{3} $ in equation (\ref{3.1}) is reduced as a result of the
substitution $z\rightarrow z+c_{3}$, therefore, without loss of generality,
we further set $c_{3}=0$.{\Huge \ }It follows from Eq. (\ref{3.7}) that%
\begin{equation}
	2\mathrm{Re}\left( g^{\ast }\left( \tau \right) f\left( \tau \right) \right) =2%
	\mathrm{Re}\left( c_{1}^{\ast }c_{2}\right) =1\Longrightarrow \left\vert
	c_{2}\right\vert \left\vert c_{1}\right\vert \cos \left( \mu _{2}-\mu
	_{1}\right) =\frac{1}{2}\ ,  \label{3.5}
\end{equation}%
where $c_{1}=\left\vert c_{1}\right\vert e^{i\mu _{1}}$ and $%
c_{2}=\left\vert c_{2}\right\vert e^{i\mu _{2}}$, $\mu _{1}\in \lbrack
0;2\pi )$, $\mu _{2}\in \lbrack 0;2\pi )$. Taking all that into account, we
obtain:%
\begin{align}
	& q=g^{\ast }\left( \tau \right) \hat{A}\left( \tau \right) +g\left( \tau
	\right) \hat{A}^{\dag }\left( \tau \right) -2\mathrm{Re}\left( g^{\ast }(\tau
	)\varphi \left( \tau \right) \right) \ ,  \notag \\
	& i\hat{p}_{q}=c_{1}^{\ast }\hat{A}\left( \tau \right) -c_{1}\hat{A}^{\dag
	}\left( \tau \right) -2i\mathrm{Im}\left( c_{1}^{\ast }\varphi \left( \tau
	\right) \right) \ .  \label{3.8}
\end{align}

\subsection{GCS}

Consider the eigenvalue problem $\hat{A}\left( \tau \right) \left\vert
z,\tau \right\rangle =z(\tau )\left\vert z,\tau \right\rangle $ for the
annihilation operator $\hat{A}\left( \tau \right) .$ In the general case
eigenvalues $z(\tau )$ that correspond to eigenvectors $\left\vert z,\tau
\right\rangle $ depend on the time $\tau $. However, if $\hat{A}\left( \tau
\right) $ is the integral of motion and, at the same time $\left\vert z,\tau
\right\rangle $ are normalized solutions of the corresponding Schr\"{o}%
dinger equation $\hat{S}\left\vert z,\tau \right\rangle =0$, these
eigenvalues do not depend on time. A simple proof of this statement follows
from the fact that if $\hat{A}\left( \tau \right) $ is the integral of
motion its mean value in the state $\left\vert z,\tau \right\rangle $ does
not depend on time. Thus,%
\begin{equation*}
	\left\langle z,\tau \left\vert \hat{A}\left( \tau \right) \right\vert z,\tau
	\right\rangle =z(\tau )=\mathrm{const\ }=z\ .
\end{equation*}%
A more formal proof, based on the equations $\hat{S}\left\vert z,\tau
\right\rangle =0$ and $\left\vert z,\tau \right\rangle \neq 0$ is given by a
chain of relations:%
\begin{align}
	\left[ \hat{S},\hat{A}\left( \tau \right) \right] \left\vert z,\tau
	\right\rangle & =\hat{S}\left( z(\tau )\left\vert z,\tau \right\rangle \right)\nonumber\\
	& =i\dot{z}(\tau )\left\vert z,\tau \right\rangle +z(\tau )\hat{S}\left\vert
	z,\tau \right\rangle =i\dot{z}(\tau )\left\vert z,\tau \right\rangle
	=0\Longrightarrow\nonumber\\
	z(\tau )=\mathrm{const\ }&=z\ .  \label{4.0}
\end{align}%
Thus, in what follows, the above mentioned eigenvalue problem looks as
follows:%
\begin{equation}
	\hat{A}\left(\tau\right) \left\vert z,\tau \right\rangle = 
	z\left\vert z,\tau \right\rangle,\quad
	\left\langle z,\tau \mid z,\tau \right\rangle = 1\ ,
	\label{4.1}
\end{equation}%
where in the general case $z$ is a complex number.

It follows from Eqs. (\ref{3.8}) and (\ref{4.1}) that%
\begin{align}
	& q\left( \tau \right) \equiv \left\langle z,\tau \left\vert q\right\vert
	z,\tau \right\rangle =q_{0}+p_{0}\tau +F_{q}\frac{\tau ^{2}}{2},\ \ q_{0}=2%
	\mathrm{Re}\left( c_{2}^{\ast }z\right) \ ,  \notag \\
	& p\left( \tau \right) \equiv \left\langle z,\tau \left\vert \hat{p}%
	\right\vert z,\tau \right\rangle =p_{0}+F_{q}\tau ,\ \ p_{0}=2\mathrm{Im}%
	\left( c_{1}^{\ast }z\right) \ ,  \notag \\
	& z=\left\langle z,\tau \left\vert \hat{A}\left( \tau \right) \right\vert
	z,\tau \right\rangle =c_{1}q\left( \tau \right) +ig\left( \tau \right)
	p(\tau )+\varphi \left( \tau \right) =c_{1}q_{0}+ic_{2}p_{0}\ .  \label{4.2}
\end{align}%
The mean values $q\left( \tau \right) $ and $p\left( \tau \right) $
correspond to the classical trajectory of accelerated by a constant force $%
F_{q}$ particle. They satisfy the classical Hamilton equations with the
Hamiltonian $H_{q}$.

Being written in the coordinate representation, equation (\ref{4.1}) reads:%
\begin{equation}
	\left[ c_{1}q+\varphi \left( \tau \right) +g\left( \tau \right) \partial _{q}%
	\right] \Phi _{z}\left( q,\tau \right) =z\Phi _{z}\left( q,\tau \right) ,\ \
	\Phi_{z}\left( q,\tau \right) \equiv \left\langle q\mid z,\tau \right\rangle \ .
	\label{4.3}
\end{equation}%
The general solution of this equation has the form%
\begin{equation}
	\Phi _{z}\left( q,\tau \right) =N\exp \left[ -\frac{c_{1}}{g\left( \tau
		\right) }\frac{q^{2}}{2}+\frac{z-\varphi \left( \tau \right) }{g\left( \tau
		\right) }q+\chi \left( \tau ,z\right) \right] \ ,  \label{4.4}
\end{equation}%
where $\chi \left( \tau ,z\right) $ is an arbitrary function on $\tau $ and $%
z$ and $N$ is a normalization constant.

One can see that the functions $\Phi _{z}\left( q,\tau \right) $ can be
written in terms of the mean values $q\left( \tau \right) $ and $p\left(
\tau \right) $,%
\begin{equation}
	\Phi _{z}\left( q,\tau \right) =N\exp \left\{ ip(\tau )q-\frac{c_{1}}{%
		2g\left( \tau \right) }\left[ q-q\left( \tau \right) \right] ^{2}+\phi
	\left( \tau ,z\right) \right\} \ ,  \label{4.5}
\end{equation}%
where $\phi \left( \tau ,z\right) $ is an arbitrary function on $\tau $ and $%
z$.

We now demand the functions $\Phi _{z}\left( q,\tau \right) $ satisfy the
Schr\"{o}dinger equation%
\begin{equation}
	\hat{S}\Phi _{z}\left( q,\tau \right) =0\ ,  \label{4.6}
\end{equation}%
where the operator $\hat{S}$ is defined in Eq. (\ref{1.15}). Thus, we fix the
function $\phi \left( \tau ,z\right) $ to be:%
\begin{equation}
	\phi \left( \tau ,z\right) =-\frac{i}{2}\int_{0}^{\tau }\left[ p^{2}(\tau
	^{\prime })+\frac{f(\tau ^{\prime })}{g(\tau ^{\prime })}\right] d\tau
	^{\prime }\ .  \label{4.7a}
\end{equation}

The density probability $\rho \left( q,\tau \right) $ generated by the
functions $\Phi _{z}\left( q,\tau \right) $ has the form:%
\begin{equation}
	\rho \left( q,\tau \right) =\left\vert \Phi _{z}\left( q,\tau \right)
	\right\vert ^{2}=N^{2}\exp \left\{ -\frac{\left[ q-q\left( \tau \right) %
		\right] ^{2}}{2\left\vert g\left( \tau \right) \right\vert ^{2}}+2\mathrm{Re}%
	\phi \left( \tau ,z\right) \right\} \ .  \label{4.8}
\end{equation}%
Considering the normalization integral, we find the constant $N$,%
\begin{equation}
	\int_{-\infty }^{\infty }\rho \left( q,\tau \right) dq=1\Rightarrow N=\left(
	2\pi \left\vert g\left( \tau \right) \right\vert ^{2}\right) ^{-1/4}\exp
	\left( -\mathrm{Re}\phi \left( \tau ,z\right) \right) \ .  \label{4.9}
\end{equation}%
Thus, normalized solutions of the Schr\"{o}dinger equation that are
eigenfunctions of the annihilation operator $\hat{A}\left( \tau \right) $
have the form:%
\begin{equation}
	\Phi _{z}\left( q,\tau \right) =\frac{1}{\sqrt{\sqrt{2\pi }\left\vert
			g\left( \tau \right) \right\vert }}\exp \left\{ ip(\tau )q-\frac{f(\tau )}{%
		g\left( \tau \right) }\frac{\left[ q-q\left( \tau \right) \right] ^{2}}{2}+i%
	\mathrm{Im}\phi \left( \tau ,z\right) \right\} \ .  \label{4.10a}
\end{equation}%
whereas the corresponding probability density reads:%
\begin{equation}
	\rho _{z}\left( q,\tau \right) =\frac{1}{\sqrt{2\pi }\left\vert g\left( \tau
		\right) \right\vert }\exp \left\{ -\frac{\left[ q-q\left( \tau \right) %
		\right] ^{2}}{2\left\vert g\left( \tau \right) \right\vert ^{2}}\right\} \ .
	\label{4.11}
\end{equation}

In what follows we call solutions (\ref{4.10a}) the time-dependent
generalized CS. It should be noted that, in fact, we have a family of states
parametrized by two complex constants $c_{1}$ and $c_{2}$ that satisfy
restriction (\ref{3.5}). Additional restrictions on the constants $c_{1}$
and $c_{2}$ transform these states into CS of the accelerated particle, see
below.

Substituting the explicit form of trajectories (\ref{4.2}) into Eq. (\ref%
{4.7a}), obtain the function $\phi \left( \tau ,z\right) $ in the following
form: 
\begin{equation}
	\phi \left( \tau ,z\right) =-\frac{i}{2}\left( F_{q}^{2}\frac{\tau ^{3}}{3}%
	+F_{q}p_{0}\tau ^{2}+p_{0}^{2}\tau \right) -\frac{1}{2}\ln \frac{g\left(
		\tau \right) }{c_{2}}\ .  \label{4.7b}
\end{equation}%
Thus we obtain a general formula for GCS of an accelerated particle,%
\begin{align}
	& \Phi _{z}\left( q,\tau \right) =\frac{1}{\sqrt{\sqrt{2\pi }\frac{\left\vert
				c_{2}\right\vert }{c_{2}}g\left( \tau \right) }}\nonumber\\
	& \times\exp \left\{ i\left[ p(\tau
	)q-\frac{1}{2}p_{0}^{2}\tau \right] -\frac{c_{1}}{g\left( \tau \right) }%
	\frac{\left[ q-q\left( \tau \right) \right] ^{2}}{2}-\frac{i}{2}F_{q}\left( 
	\frac{F_{q}}{3}\tau +p_{0}\right) \tau ^{2}\right\} \ .  \label{4.10}
\end{align}

Setting $F_{q}=0$ in Eq. (\ref{4.10}), we obtain the time-dependent
generalized CS of a free particle, see Ref. \cite{303}.

Next we will demonstrate that GCS satisfy the completeness condition. To
this end we first consider the action of the displacement operator $\mathcal{%
	D}\left( z,\tau \right) =\exp \left[ z\hat{A}^{\dag }\left( \tau \right)
-z^{\ast }\hat{A}\left( \tau \right) \right] $ on the vacuum vector $%
\left\vert 0,\tau \right\rangle $ in the coordinate representation:%
\begin{eqnarray}
	\tilde{\Phi}_{z}\left( q,\tau \right) &=&\mathcal{D}\left( z,\tau \right)
	\Phi _{0}\left( q,\tau \right) =\exp \left[ -\frac{\left\vert z\right\vert
		^{2}}{2}+z\hat{A}^{\dag }\left( \tau \right) \right] \Phi _{0}\left( q,\tau
	\right) ,  \label{4.12} \\
	\ \Phi _{0}\left( q,\tau \right) &=&\left\langle q\mid 0,\tau \right\rangle =%
	\frac{1}{\sqrt{\sqrt{2\pi }\frac{\left\vert c_{2}\right\vert }{c_{2}}g\left(
			\tau \right) }}\nonumber\\
	&\times&\exp \left[ -\frac{c_{1}}{g\left( \tau \right) }\frac{\left(
		q-F_{q}\frac{\tau ^{2}}{2}\right) ^{2}}{2}+iF_{q}\left( q-F_{q}\frac{\tau
		^{2}}{6}\right) \tau \right] .  \notag
\end{eqnarray}%
Thus, taking the explicit forms of the mean values (\ref{4.2}) into account,
we obtain:%
\begin{align}
	\tilde{\Phi}_{z}\left( q,\tau \right) &=\exp \left\{ -\frac{\left\vert
		z\right\vert ^{2}}{2}+\left[ c_{1}^{\ast }\left( q-g^{\ast }(\tau )\frac{z}{2%
	}\right) +\varphi \left( \tau \right) \right] z\right\} \Phi _{0}\left(
	q-g^{\ast }(\tau )z,\tau \right)\nonumber\\
	& =\exp \left( -\frac{i}{2}q_{0}p_{0}\right)
	\Phi _{z}\left( q,\tau \right) \ .  \label{4.14}
\end{align}%
The states $\Phi _{z}\left( q,\tau \right) $ and$\ \tilde{\Phi}_{z}\left(
q,\tau \right) $ differ by a phase factor only.

Now we will show that the states $\tilde{\Phi}_{z}\left( q,\tau \right) $
satisfy the completeness condition, which will give us the completeness
condition for GCS. To this end, It is useful to introduce the vacuum vector $%
\left\vert 0,\tau \right\rangle $ at a given time instant, $\hat{A}\left(
\tau \right) \left\vert 0,\tau \right\rangle =0$, and the corresponding Fock
space,%
\begin{eqnarray}
	&&\left\vert n,\tau \right\rangle =\frac{\left[ \hat{A}^{\dag }\left( \tau
		\right) \right] ^{n}}{\sqrt{n!}}\left\vert 0,\tau \right\rangle ,\ \ \langle
	n,\tau \left\vert n,\tau \right\rangle =1\ ,\ \ n=0,1,2,\dots \ ,  \notag \\
	&&\hat{A}\left( \tau \right) \left\vert n,\tau \right\rangle =\sqrt{n}%
	\left\vert n-1,\tau \right\rangle ,\ \ \hat{A}^{\dag }\left( \tau \right)
	\left\vert n,\tau \right\rangle =\sqrt{n+1}\left\vert n+1,\tau \right\rangle
	\ .  \label{3.9}
\end{eqnarray}%
Using representation (\ref{4.12}) and definitions (\ref{3.9}), one derives
the following form for the states $\tilde{\Phi}_{z}\left( q,\tau \right) $:%
\begin{equation}
	\tilde{\Phi}_{z}\left( q,\tau \right) =\exp \left[ -\frac{\left\vert
		z\right\vert ^{2}}{2}\right] \sum_{n=0}^{\infty }\frac{z^{n}}{\sqrt{n!}}%
	\langle q\left\vert n,\tau \right\rangle \ .  \label{4.15}
\end{equation}%
With the help the completeness property of the states $\left\vert n,\tau
\right\rangle $,%
\begin{equation}
	\sum_{n=0}^{\infty }\left\vert n,\tau \right\rangle \left\langle n,\tau
	\right\vert =1,\ \ \forall \tau \ ,  \label{4.17}
\end{equation}%
one can find the overlapping of the CS and prove the corresponding
completeness relations:%
\begin{align}
	& \int_{-\infty }^{+\infty }\left( \tilde{\Phi}_{z^{\prime }}\left( q,\tau
	\right) \right) ^{\ast }\tilde{\Phi}_{z}\left( q,\tau \right) dq=\exp \left(
	z^{\prime \ast }z-\frac{\left\vert z^{\prime }\right\vert ^{2}+\left\vert
		z\right\vert ^{2}}{2}\right) ,\ \ \forall \tau \ ;  \notag \\
	& \int \int \left( \tilde{\Phi}_{z}\left( q,\tau \right) \right) ^{\ast }%
	\tilde{\Phi}_{z}\left( q^{\prime },\tau \right) d^{2}z=\pi \delta \left(
	q-q^{\prime }\right) d^{2}z=d\mathrm{Re}z d\mathrm{Im}z,\text{ \ }%
	\forall \tau \ .  \label{4.18}
\end{align}%
Eqs. (\ref{4.18}) imply already the completeness relation for the GCS,%
\begin{equation*}
	\int \int \left( \Phi _{z}\left( q,\tau \right) \right) ^{\ast }\Phi
	_{z}\left( q^{\prime },\tau \right) d^{2}z=\pi \delta \left( q-q^{\prime
	}\right) ,\text{ \ }\forall \tau \ .
\end{equation*}

\section{Standard deviations and conditions of semi-classicality\label{S4}}

Calculating standard deviations $\sigma _{q}\left( \tau \right) $, $\sigma
_{p}\left( \tau \right) $, and the characteristic quantity $\sigma
_{qp}\left( \tau \right) $, with respect to the GCS, we obtain:%
\begin{align}
	\sigma _{q}\left( \tau \right) &=\sqrt{\langle \left( \hat{q}-\left\langle
		q\right\rangle \right) ^{2}\rangle }=\sqrt{\left\langle q^{2}\right\rangle
		-\left\langle q\right\rangle ^{2}}=\left\vert g\left( \tau \right)
	\right\vert \ ,  \notag \\
	\sigma _{p}\left( \tau \right) &=\sqrt{\langle \left( \hat{p}-\left\langle
		p\right\rangle \right) ^{2}\rangle }=\sqrt{\left\langle p^{2}\right\rangle
		-\left\langle p\right\rangle ^{2}}=\left\vert f\left( \tau \right)
	\right\vert =\left\vert c_{1}\right\vert \ ,  \notag \\
	\sigma _{qp}\left( \tau \right) &=\frac{1}{2}\left\langle \left( \hat{q}%
	-\left\langle q\right\rangle \right) \left( \hat{p}-\left\langle
	p\right\rangle \right) +\left( \hat{p}-\left\langle p\right\rangle \right)
	\left( \hat{q}-\left\langle q\right\rangle \right) \right\rangle\nonumber\\
	&=i\left[1/2-g\left( \tau \right) f^{\ast }\left( \tau \right) \right] \ .
	\label{4.19}
\end{align}%
One can easily see that the GCS, for any set of the parameters $c_{1}$ and $%
c_{2}$,\ minimize the Robertson-Schr\"{o}dinger uncertainty relation \cite%
{SchroRo,SchroRo2},%
\begin{equation}
	\sigma _{q}^{2}\left( \tau \right) \sigma _{p}^{2}(\tau )-\sigma
	_{qp}^{2}\left( \tau \right) =1/4\ .  \label{4.20}
\end{equation}

Let us consider the Heisenberg uncertainty relation for the GCS. Taking into
account relations (\ref{3.5}) for constants $c_{1}$ and $c_{2}$, we obtain%
\begin{equation}
	\sigma _{q}\left( \tau \right) \sigma _{p}\left( \tau \right) =\sqrt{\frac{1%
		}{4}+\left[ \left\vert c_{2}\right\vert \left\vert c_{1}\right\vert \sin
		\left( \mu _{2}-\mu _{1}\right) +\left\vert c_{1}\right\vert ^{2}\tau \right]
		^{2}}\geq \frac{1}{2}\ .  \label{4.21}
\end{equation}%
Then using Eq. (\ref{4.19}), we find $\sigma _{q}\left( 0\right) =\sigma
_{q}=\left\vert c_{2}\right\vert $ and $\sigma _{p}\left( 0\right) =\sigma
_{p}=\left\vert c_{1}\right\vert $, such that at $\tau =0$ the Eq. (\ref%
{4.21}) implies: 
\begin{equation}
	\sigma _{q}\sigma _{p}=\left\vert c_{2}\right\vert \left\vert
	c_{1}\right\vert =\sqrt{\frac{1}{4}+\left[ \left\vert c_{2}\right\vert
		\left\vert c_{1}\right\vert \sin \left( \mu _{2}-\mu _{1}\right) \right] ^{2}%
	}\geq \frac{1}{2}\ .  \label{4.22}
\end{equation}

It follows from (\ref{3.5}) that $\left\vert c_{i}\right\vert \neq 0$, $i=1$%
, $2$ say that the left hand side of Eq. (\ref{4.22}) is minimized when $\mu
_{1}=\mu _{2}=\mu $, which provides the minimization of the Heisenberg
uncertainty relation for the CS at the initial time instant,%
\begin{equation}
	\left. \sigma _{q}\left( \tau \right) \sigma _{p}\left( \tau \right)
	\right\vert _{\tau =0}=\frac{1}{2}\ .  \label{4.23}
\end{equation}

In what follows, we consider the GCS with the above restriction $\mu
_{1}=\mu _{2}=\mu $. Namely, such states we call simply CS of a accelerated
particle. In this case relation (\ref{3.5}), $2\mathrm{Re}\left( c_{1}^{\ast
}c_{2}\right) =1$, takes the form:%
\begin{equation}
	\left\vert c_{2}\right\vert \left\vert c_{1}\right\vert =1/2\Longrightarrow
	c_{2}^{\ast }=c_{1}^{-1}/2\ .  \label{4.24}
\end{equation}%
One can see that the constant $\mu $ does not enter the CS. Thus, in what
follows we set $\mu =0$. Then%
\begin{align}
	& c_{2}=\left\vert c_{2}\right\vert =\sigma _{q},\ c_{1}=\left\vert
	c_{1}\right\vert =\sigma _{p}=1/(2\sigma _{q})\ ,  \notag \\
	& g\left( \tau \right) =\left( \sigma _{q}+\frac{i\tau }{2\sigma _{q}}%
	\right) ,\ \ \sigma _{q}\left( \tau \right) =\left\vert g\left( \tau \right)
	\right\vert =\sqrt{\sigma _{q}^{2}+\frac{\tau ^{2}}{4\sigma _{q}^{2}}}\ .
	\label{4.25}
\end{align}%
It follows from Eqs. (\ref{4.25}), that for any time instant $\tau $ the
Heisenberg uncertainty relation for the CS takes the form: 
\begin{equation}
	\sigma _{q}\left( \tau \right) \sigma _{p}\left( \tau \right) =\frac{1}{2}%
	\sqrt{1+\frac{\tau ^{2}}{4\sigma _{q}^{4}}}\geq \frac{1}{2}\ .  \label{4.26}
\end{equation}

Finally, taking into account Eqs. (\ref{4.2}), we obtain the following
coordinate representation for the CS of an accelerated particle:%
\begin{equation}
	\Phi _{z}^{\sigma _{q}}\left( q,\tau \right) =\frac{\exp \left\{ i\left[
		p(\tau )q-\frac{p_{0}^{2}}{2}\tau \right] -\frac{\left[ q-q\left( \tau
			\right) \right] ^{2}}{4\left( \sigma _{q}^{2}+i\tau /2\right) }-\frac{i}{2}%
		F_{q}\left( F_{q}\frac{\tau }{3}+p_{0}\right) \tau ^{2}\right\} }{\sqrt{%
			\left( \sigma _{q}+\frac{i\tau }{2\sigma _{q}}\right) \sqrt{2\pi }}}\ .
	\label{4.27}
\end{equation}%
\ 

We stress that, in fact, we have constructed a family of the CS parametrized
by one real parameter $\sigma _{q}$. Each set of the CS in the family has
its specific initial standard deviations $\sigma _{q}>0$. The CS from a
family with a given $\sigma _{q}$ can be also labeled by the quantum number $%
z$,%
\begin{equation}
	z=\frac{q_{0}}{2\sigma _{q}}+i\sigma _{q}p_{0}\ ,  \label{4.28}
\end{equation}%
which is in one to one correspondence with the corresponding classical
trajectory initial data, $q_{0}=2\sigma _{q}\mathrm{Re}z$,\ $\ p_{0}=(\mathrm{Im}%
z)/\sigma _{q}$. Thus, we will take $\sigma _{q}$ and $z$ as independent
parameters of the constructed CS.

If $\sigma _{q}<1/2$ or $\sigma _{p}<1/2$ the accelerated particle CS are,
at the initial time instant, the so-called squeezed states; see Ref. \cite{DodMa03}.

The probability densities that corresponds to the CS (\ref{4.27}) are:%
\begin{equation}
	\rho _{z}^{\sigma _{q}}\left( q,\tau \right) =\frac{1}{\sqrt{2\pi \sigma
			_{q}^{2}\left( \tau \right) }}\exp \left\{ -\frac{\left[ q-q\left( \tau
		\right) \right] ^{2}}{2\sigma _{q}^{2}\left( \tau \right) }\right\} \ .
	\label{4.30}
\end{equation}%
One can see that at any time instant $\tau $ the probability densities (\ref%
{4.30}) are given by Gaussian distributions with standard deviations $\sigma
_{q}\left( \tau \right) $.

Let us consider the shape of the particle wave packet (the shape of the
probability density) at the initial time instant. Eq. (\ref{4.30}) implies
that this packet has the height $L=1/(\sqrt{2\pi }\sigma _{q})$ and the
half-width $\Delta l=\sqrt{8\ln 2}\sigma _{q}$,%
\begin{equation}
	\Delta l=\frac{1}{L}\sqrt{\frac{4\ln 2}{\pi }}\approx 0.939\frac{1}{L}\ .
	\label{4.31}
\end{equation}%
The same relation holds true for the all the GCS.

Fig. 1 shows the wave packet corresponding to the CS (\ref{4.27}) at the
initial time for $\sigma _{q}=0.2$ and $q_{0}=0$.

\begin{figure}[h]
	\centering
	\includegraphics[width=0.4\textwidth]{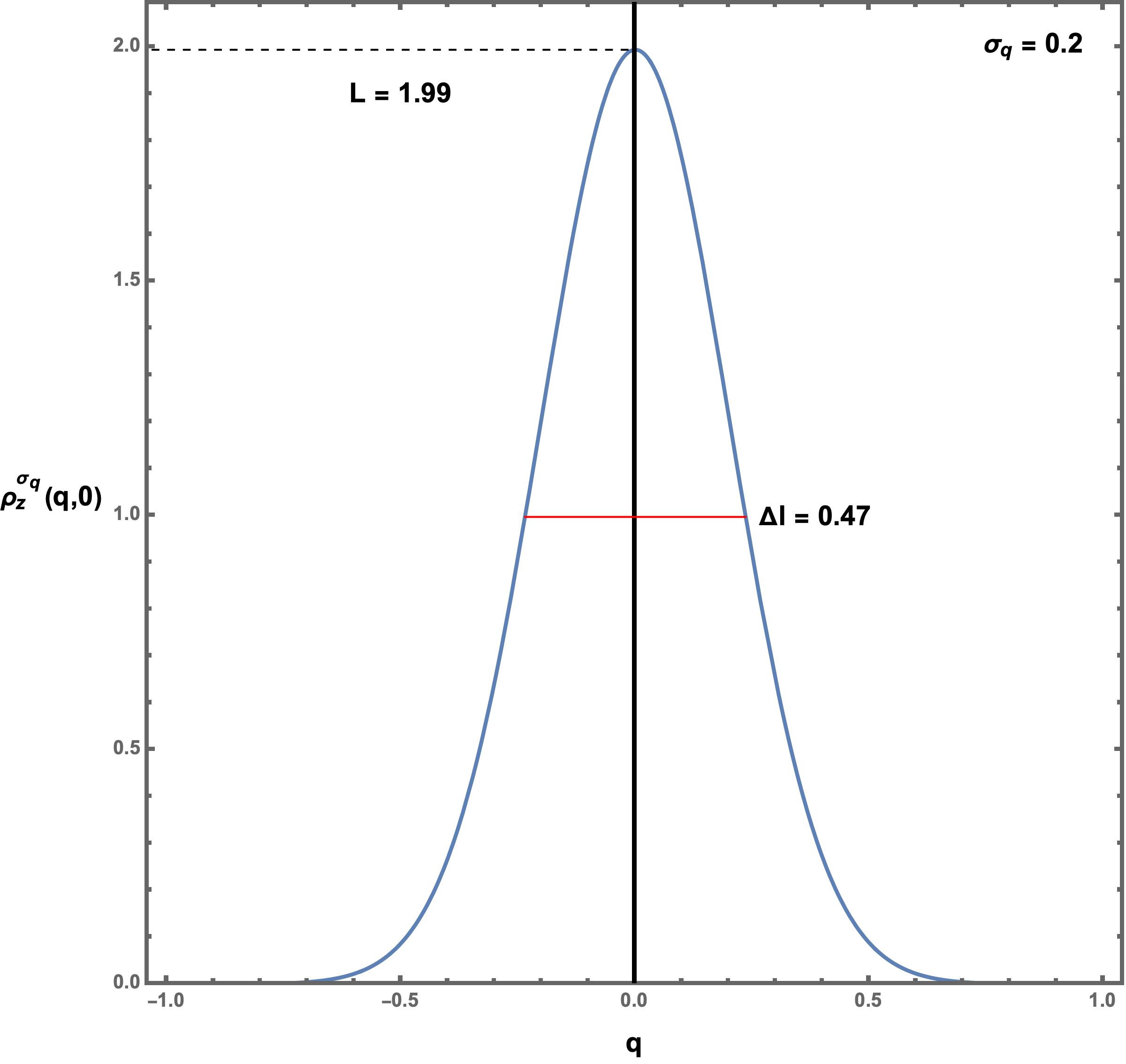}
	\caption{The shape of the wave packet at the initial moment of time for $\protect\sigma _{q}=0.2$ and $q_{0}=0$.}
	\label{Fig1}
\end{figure}

Now consider the change in the shape of the wave packet over the time. The
coordinate mean values $\left\langle q\right\rangle =q\left( \tau \right)
=q_{0}+p_{0}\tau +F_{q}\tau ^{2}/2$ are moving along the classical
trajectory with the particle momentum $\left\langle p\right\rangle =p\left(
\tau \right) =p_{0}+F_{q}\tau $ and the constant acceleration $F_{q}$. With
the same momentum and the acceleration are moving the maxima of the
probability densities (\ref{4.30}). The half-width $\Delta l(\tau )=\sqrt{%
	8\ln 2}\sigma _{q}(\tau )$ of a given Gaussian wave packet does not depend
on the force $F$ acting on the particle. This force affects the magnitude of
the shift of the wave packet as a whole along the coordinate axis $q$ per
time unit.

\begin{figure}[h]
	\centering
	\includegraphics[width=0.5\textwidth]{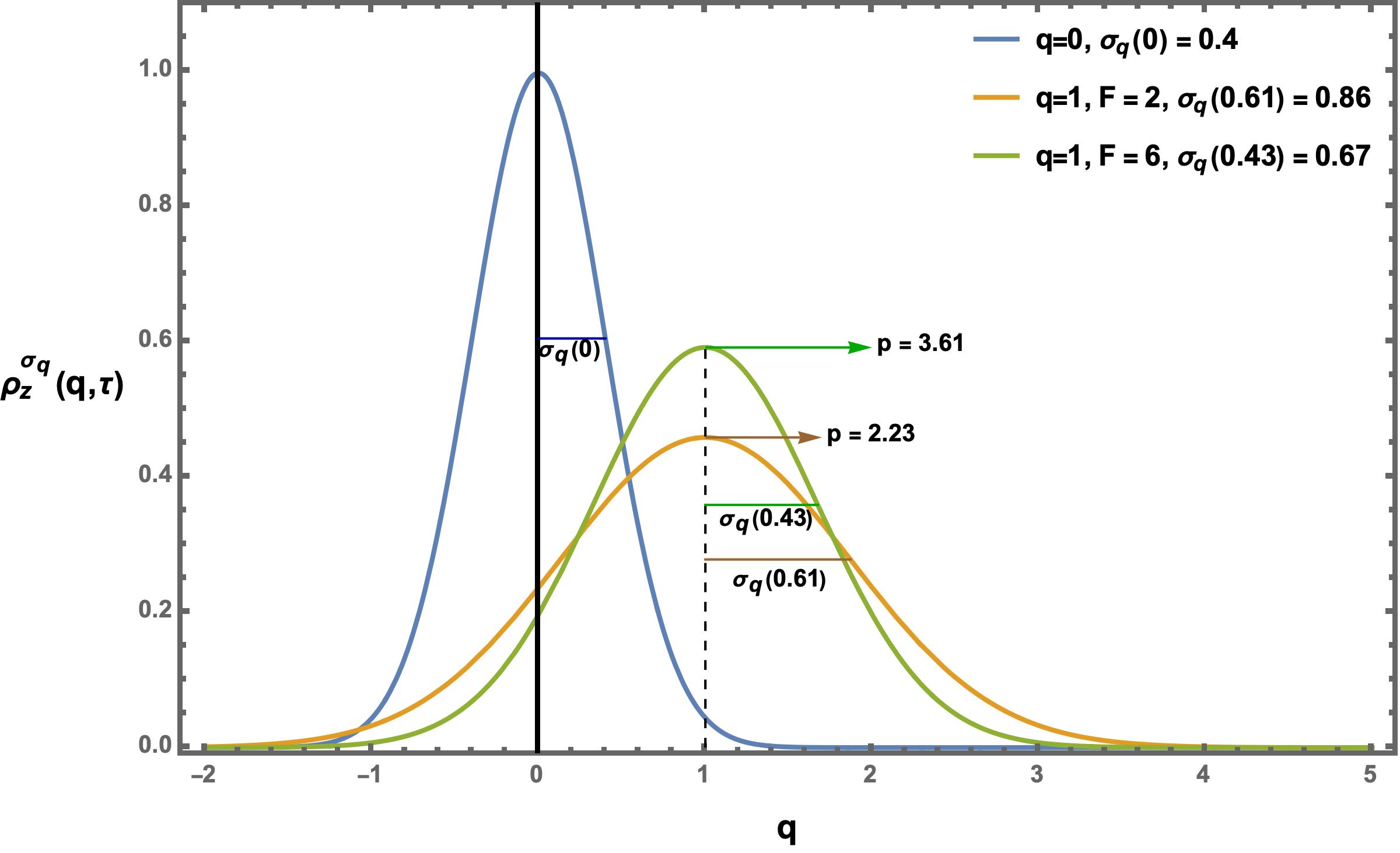}
	\caption{Evolution of the probability density for $q_{0}=0$, $p_{0}=1$, 
		$\sigma_{q}=0.4$. Blue color shows the particle distribution density at the initial moment of time with initial conditions $q_{0}=0$, $p_{0}=1$. The standard deviation at the initial time is chosen to be $\sigma _{q}=0.4$. Yellow color shows the evolution of the distribution density of an accelerated particle, which is acted upon by a force $F=2$ and which is located at point $q=1$ and has momentum $p=2.23$. Green color shows the evolution of the distribution density of an accelerated particle at the same point $q=1$ with momentum $p=3.61$, which is acted upon by a force $F=6$.}
	\label{Fig2}
\end{figure}

The maximum of the probability density (\ref{4.30}) is located at the point $%
q>q_{0}$ at the time%
\begin{equation}
	\tau =\tau _{q}=\left\{ 
	\begin{array}{c}
		\left[ \sqrt{\left( \frac{p_{0}}{F_{q}}\right) ^{2}+\frac{2(q-q_{0})}{F_{q}}}%
		-\frac{p_{0}}{F_{q}}\right] ,\ \ F_{q}>0 \\ 
		(q-q_{0})/p_{0},\ \ F_{q}=0%
	\end{array}%
	\right. \ ,  \label{4.32}
\end{equation}%
and is characterized by the standard deviation $\Omega _{q}$,%
\begin{eqnarray}
	\Omega _{q} &&=\sigma _{q}(\tau _{q})=\sqrt{\sigma _{q}^{2}+\frac{1}{4\sigma
			_{q}^{2}}\left[ \sqrt{\left( \frac{p_{0}}{F_{q}}\right) ^{2}+\frac{2(q-q_{0})%
			}{F_{q}}}-\frac{p_{0}}{F_{q}}\right] ^{2}}  \notag \\
	&&\ <\left. \Omega _{q}\right\vert _{F_{q}=0}=\sqrt{\sigma _{q}^{2}+\left( 
		\frac{q-q_{0}}{2p_{0}\sigma _{q}}\right) ^{2}}\ .  \label{4.33}
\end{eqnarray}

The spreading of the wave packet of an accelerated particle at a point $q$
is less than the spreading of the wave packet of a free particle arriving at
the same point. This blurring decreases the larger the $F$:%
\begin{equation}
	\Omega _{q}=\sigma _{q}+\frac{(q-q_{0})}{4F_{q}\sigma _{q}^{3}}+O\left( 
	\frac{1}{F_{q}}\right) ^{3/2}\ .  \label{4.34}
\end{equation}%
Let us illustrate what has been said with the graph Fig. 2. 
We see that the greater the force $F$, the less the spreading of the wave packet corresponding to the particle at point $q=1$.

To consider the question which CS\ can be treated as representing a
semiclassical particle motion, we have to return to the initial dimensional
variables $x$ and $t$ (\ref{1.12c}) and to the initial wave function $\Psi_z^{\sigma_q}
\left( x,t\right) $ written in these variables,
\begin{equation}
	\Phi _{z}^{\sigma _{q}}\left( q,\tau \right) =\sqrt{l}\Psi_z^{\sigma _{q}} \left( lq,\frac{ml^{2}}{\hbar }%
	\tau \right)\,.\nonumber
\end{equation} 
Taking into account that
\begin{align}
	& x\left( t\right) =lq\left( \tau \right) =x_{0}+\frac{p_{0}^{x}}{m}t+\frac{%
		F_{x}}{m}\frac{t^{2}}{2},\ \ p_{0}=\frac{l}{\hbar }p_{0}^{x}\ ,  \notag \\
	& p_{x}(t)=\frac{\hbar }{l}p\left( \tau \right) =p_{0}^{x}+F_{x}t\ ,  \notag
	\\
	& \sigma _{x}\left( 0\right) =l\sigma _{q}\left( 0\right) =l\sigma =\sigma
	_{x},\ \sigma _{x}^{2}\left( t\right) =\sigma _{x}^{2}+\frac{\hbar ^{2}}{%
		4m^{2}\sigma _{x}^{2}}t^{2}\ ,  \label{e18a}
\end{align}%
we obtain%
\begin{align}
	& \Psi_z^{\sigma_q} \left( x,t\right) =\frac{1}{\sqrt{\left( \sigma _{x}+\frac{i\hbar }{%
				2m\sigma _{x}}t\right) \sqrt{2\pi }}}\nonumber\\
	&\times\exp \left\{ \frac{i}{\hbar }\left[
	\left( p_{x}(\tau )x-\frac{p_{0}^{x2}}{2m}t\right) -\frac{F_{x}}{m}\left( 
	\frac{F_{x}}{3}t+p_{0}^{x}\right) \frac{t^{2}}{2}\right] -\frac{\left[
		x-x\left( t\right) \right] ^{2}}{4\left( \sigma _{x}^{2}+\frac{\hbar }{2m}%
		it\right) }\right\} \ ,  \notag \\
	& \rho_z^{\sigma_q} \left( x,t\right)=\nonumber\\
	& =\left\vert \Psi_z^{\sigma_q} \left( x,t\right) \right\vert ^{2}=%
	\frac{1}{\sqrt{\left( \sigma _{x}^{2}+\frac{\hbar ^{2}}{4m^{2}\sigma _{x}^{2}%
			}t^{2}\right) 2\pi }}\exp \left\{ -\frac{1}{2}\frac{\left[ x-x\left(
		t\right) \right] ^{2}}{\sigma _{x}^{2}+\frac{\hbar ^{2}}{4m^{2}\sigma
			_{x}^{2}}t^{2}}\right\} \ .  \label{e20}
\end{align}

The shape of distribution (\ref{e20}) that corresponds to the semiclassical
motion must change with the time slowly in a certain sense. This change is
associated with a change of the quantity $\sigma _{x}^{2}\left( t\right) $,
see Eq. (\ref{e18a}). We suppose that the semiclassical motion implies the
quantity $\sigma _{x}^{2}\left( t\right) $ (or equivalently $\frac{\hbar ^{2}%
}{4m^{2}\sigma _{x}^{2}}t^{2}$) is much less than the distance square that
the particle passes for the same time $t$. Then, the semi-classicality
condition reads:%
\begin{equation}
	\frac{\hbar ^{2}t^{2}}{4\sigma _{x}^{2}}\ll \left( p_{0}^{x}t+\frac{%
		F_{x}t^{2}}{2}\right) ^{2}\ .  \label{e21a}
\end{equation}%
It can be rewritten in a different form:%
\begin{equation}
	\frac{\lambda }{\left\vert 1+\frac{\lambda }{2\pi \hbar }F_{x}\frac{t}{2}%
		\right\vert }\ll 4\pi \sigma _{x},\ \ \lambda =\frac{2\pi \hbar }{p_{x}}\ ,
	\label{e22}
\end{equation}%
where $\lambda $ is the Compton wavelength of the particle.

Thus, CS of a free particle ($F_{x}=0$) can be considered as semiclassical
states if the Compton wavelength of the particle is much less than the
coordinate standard deviation $\sigma _{x}$ at the initial time moment, see
Ref. \cite{303}. However, if $F_{x}\neq 0$ and after a sufficiently long
time period, CS of an accelerated particle can be always considered as
semiclassical ones.

\section*{Acknowledgments}

D.M.G. thanks CNPq for permanent support.

\section{Concluding remarks\label{S5}}

We study quantum states of the accelerated particle both known and new ones
obtained by us using the method of non-commutative integration of linear
differential equations. A complete set of non-stationary states (\ref{n4})
for the accelerated particle is obtained. This set is expressed via
elementary functions and is characterized by a continuous real parameter $%
\eta$, which corresponds to the initial momentum of the particle. A
connection is obtained between these solutions and stationary states, which
are determined by the Airy function (\ref{n10}).

We solved the problem of constructing GCS, in particular, semiclassical
states describing the accelerated particle, within the framework of the
consistent method of integrals of motion. We have found different
representations, coordinate one and in a Fock space, analyzing in detail all
the parameters entering in these representations. We prove the corresponding
completeness and orthogonality relations. Conditions for minimizing
uncertainty relations, were studied and the set of the corresponding
parameters was determined. From the GCS a family of states is isolated,
which usually is called the CS. This family of states is parameterized by
the real parameter $\sigma _{q}$, which has the meaning of the standard
deviation of the coordinate at the initial time instant. The CS minimize the
uncertainty relation (\ref{4.20}) at all the time instants and the
Heisenberg uncertainty relation (\ref{4.23}) at the initial time. The
probability density (\ref{4.30}) is given by a Gaussian distribution with
the standard deviations $\sigma _{q}(\tau)$ and the constructed CS are
wave packets that are solutions to the Schr\"{o}dinger equation for the
accelerated particle. Coordinate mean values are moving along classical
trajectories of the accelerated particles and coincide with trajectories of
the maximum of the wave packets. We prove the completeness and orthogonality
relations for the obtained GCS and CS.

Standard deviations for the GCS and CS are calculated. On this base, and
considering the change in the shape of wave packets with time, we define
general conditions of the semi-classicality and a class of CS that can be
identified with semiclassical states. As follows from this conditions, in
contrast to a free particle case, where CS can be considered as
semiclassical states if the Compton wavelength of the particle is much less
than the coordinate standard deviation $\sigma _{x}$ at the initial time
moment, see Ref. \cite{303}, after a sufficiently long time period, the CS
of the accelerated particle can be always considered as semiclassical ones.
It is interesting that this conclusion is matched with the one obtained in
Ref. \cite{Sazon23} in studying the Caldirola--Kanai model. Namely, there
were demonstrated that the force of resistance and viscous friction prevent
the spreading of a quasi-classical wave packet. Thus, the resistance force
suppresses the quantum properties of the particle, increasingly highlighting
the classical features in its movement over time.

\end{document}